\documentclass[journal]{IEEEtran}
\usepackage[T1]{fontenc}

\usepackage{multirow,graphicx}
\usepackage{amsmath}
\usepackage{color}
\usepackage{subfigure}
\usepackage{amssymb}
\usepackage{array}%
\usepackage{makecell}
\usepackage{amsmath,graphicx,multirow,algorithm,algorithmic,bm}
\usepackage[colorlinks,linkcolor=blue]{hyperref}

\ifCLASSINFOpdf
\else
\fi
\usepackage{amsmath}
\begin{document}

	%
	\title{SSCAN: A Spatial-spectral Cross Attention Network for Hyperspectral Image Denoising}
	%
	%
	%
	
	\author{Zhiqiang Wang, Zhenfeng Shao, Xiao Huang, Jiaming Wang, Tao Lu, and Sihang Zhang
		\thanks{Z. Wang is with the School of Remoter Sensing and Information Engineering, Wuhan University, Wuhan 430079, China (wangzqwhu@foxmail.com).
		
		Z. Shao, J. Wang and S. Zhang are with the State Key Laboratory for Information Engineering in Surveying, Mapping and Remote Sensing, Wuhan University, Wuhan 430079, China (e-mail: \{shaozhenfeng, wjmecho, 2019206190045\}@whu.edu.cn)
		
		X. Huang is with the Department of Geosciences, University of Arkansas, Fayetteville, AR 72701, USA (xh010@uark.edu)
		
		T. Lu is with the School of Computer Science and Engineering, Wuhan Institute of Technology, Wuhan 430205, China (e-mail: lutxyl@gmail.com).
		
		This work was supported in part by the National Key R\&D Program of China under Grant 2018YFB0505401; in part by the National Natural Science Foundation of China under Grants 41890820, 41771452, 41771454, and 62072350; in part by the Key R\&D Program of Yunnan Province in China under Grant 2018IB023. (Corresponding author: Zhenfeng Shao)}
	}

	\markboth{Journal of \LaTeX\ Class Files,~Vol.~14, No.~8, August~2015}%
	{Shell \MakeLowercase{\textit{et al.}}: Bare Demo of IEEEtran.cls for IEEE Communications Society Journals}

	\maketitle
	\begin{abstract}
	Hyperspectral images (HSIs) have been widely used in a variety of applications thanks to the rich spectral information they are able to provide. Among all HSI processing tasks, HSI denoising is a crucial step. Recently, deep learning-based image denoising methods have made great progress and achieved great performance. However, existing methods tend to ignore the correlations between adjacent spectral bands, leading to problems such as spectral distortion and blurred edges in denoisied results. In this study, we propose a novel HSI denoising network, termed SSCAN, that combines group convolutions and attention modules. Specifically, we use a group convolution with a spatial attention module to facilitate feature extraction by directing models’ attention to band-wise important features. We propose a spectral-spatial attention block (SSAB) to exploit the spatial and spectral information in hyperspectral images in an effective manner. In addition, we adopt residual learning operations with skip connections to ensure training stability. The experimental results indicate that the proposed SSCAN outperforms several state-of-the-art HIS denoising algorithms.
	\end{abstract}
	

	\IEEEpeerreviewmaketitle
	\vspace{-0.5cm}
	\section{Introduction}
	\IEEEPARstart{D}{ifferent} from traditional RGB images, hyperspectral images (HSIs) contain more spatial and spectral information that benefits a variety of remote sensing applications, such as target detection \cite{kruse2003comparison}, classifications \cite{yang2017learning} and reconnaissance \cite{bioucas2012hyperspectral}. In the last decade, the data acquisition approaches of HSI have been developed in a rapid mannerly \cite{ghamisi2017advances}. However, due to atmospheric interference, sensor restrictions, and other reasons, HISs can be corrupted by several types of noises, including Gaussian noises, stripe noises, and impulse noises \cite{2018Noise}, which largely degrade the image quality and affect subsequent applications \cite{2018Multifeature}. Therefore, denoising is considered as an essential preprocessing step for HSI applications \cite{rasti2018noise}.  
	
	HSI denoising aims to recover a clean hyperspectral image from the one with noises. 
	Early conventional methods, including block-matching 3D-filtering (BM3D) \cite{dabov2007image}, total variation (TV) \cite{rudin1992nonlinear}, and the non-local means algorithm \cite{buades2005non}, considered each spectral band as an individual gray channel, and applied 2D algorithms to each band.  Instead of treating each band separately, some studies treat HSI as a multidimensional data cube. For example, block-matching and 4D-filtering algorithms (BM4D) were applied to 3D images \cite{maggioni2012nonlocal}. Chen \emph{et al.} \cite{chen2014denoising} combined the PCA and BM4D to reduce data dimensionality. Despite their success, however, these methods largely ignore the spectral information and neglect the correlation between bands in HSIs that is considered as an effective factor in HSI denoising. To achieve better denoising performance, efforts have been made to combine spatial and spectral features based on low-rank recovery \cite{renard2008denoising,zhang2013hyperspectral,he2019non}. Renard \emph{et al.} \cite{renard2008denoising} proposed a low-rank tensor approximation method (LRTA) to reduce the dimensionality and denoise in HSI. Zhang \emph{et al.} \cite{zhang2013hyperspectral} regarded the clean HSI as a low-rank matrix and proposed low-rank matrix recovery (LRMA) for HSI denoising. He \emph{et al.}\cite{he2019non} proposed non-local meets global (NGMeet) to generate the clean HSI by fusing the spatial non-local similarity and spectral low-rank approximation. However, these traditional denoising methods are computationally demanding, which limits their practical applications. What’s more, manually selected feature extraction modules fail to exploit deep features in HSIs, leading to their unsatisfactory performances. 
	\begin{figure*}[!t]
		\centering
		\includegraphics[width=14 cm]{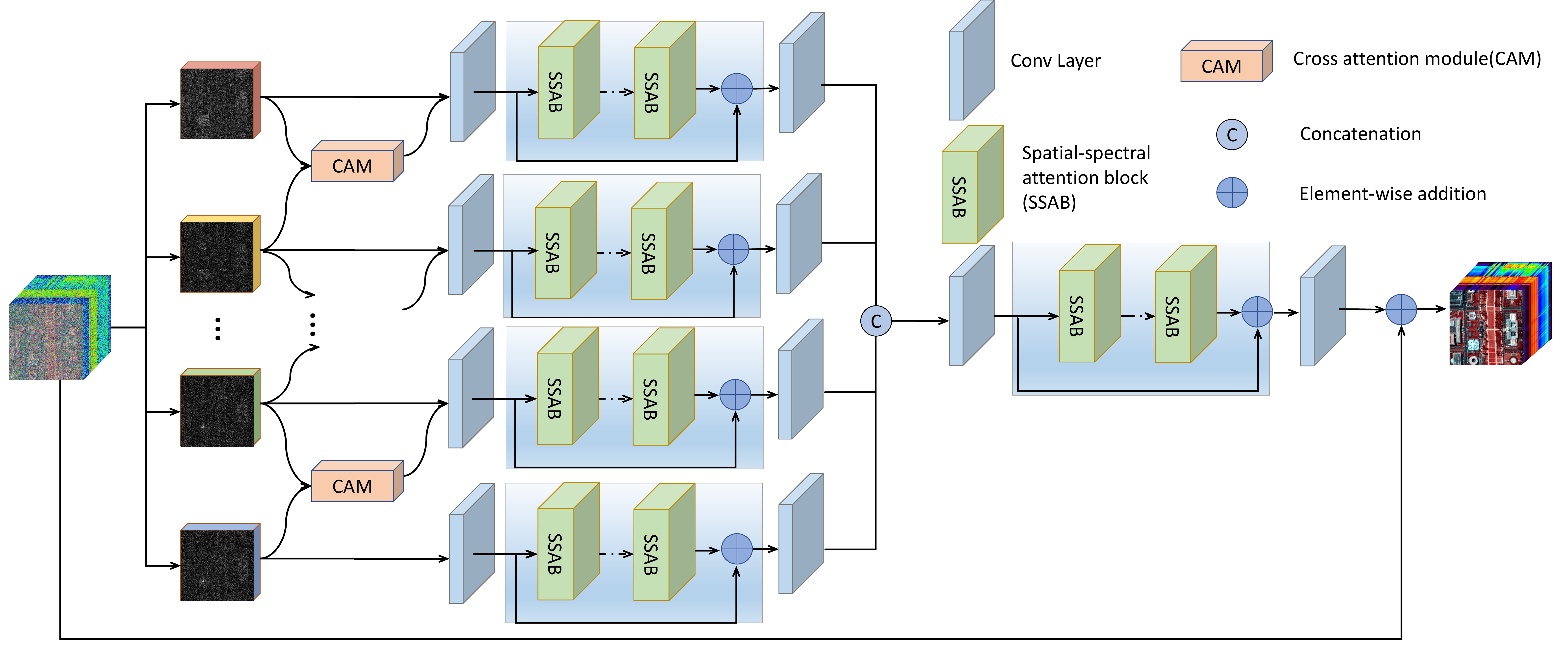}
		\caption{The overall network structure of spatial-spectral cross attention network (SSCAN), which consists of a spectral group attention module and a spatial-spectral attention network.}
		\label{fig1}
	\end{figure*}
	
	With the popularity of convolutional neural networks (CNN), deep learning-based methods exhibit great potentials in image denoising tasks. The core idea of these methods is to learn an optimal mapping function between a noisy HSI and a clean HSI, given a huge number of samples. In particular, the DnCNN proposed by Zhang \emph{et al.} \cite{zhang2017beyond} employed residual learning and batch normalization (BN) for fast convergence. RIENet \cite{anwar2019real} for the first time, applied an attention module in denoising tasks and achieved excellent performance. Further, MemNet \cite{tai2017memnet} was proposed as a deep persistent memory network to solve long-term dependency issues. Trainable non-linear reaction diffusion (TNRD) was proposed with an iterative nonlinear reaction diffusion model instead of stacking neural network layers \cite{chen2016trainable}. Following TNRD, Xie \emph{et al.} \cite{xie2017hyperspectral} migrated the same network to HSI denoising and further improved model performance. Yuan \emph{et al.} \cite{yuan2018hyperspectral} and Dong \emph{et al.} \cite{dong2019deep} combined spatial and spectral features into a network to reduce noise in HSIs, aiming to achieve better denoising performance. However, the direct migration of the methods that target natural RGB images to HSIs is not efficient, as such a migration process ignores the correlation of different spectral bands, which has a significant impact on the denoising performance in HSIs.  
	
	To take advantage of the spectral and spatial information in HSIs, while considering the correlations between adjacent bands, we propose a novel network, termed spectral-spatial cross attention network (SSCAN). In the proposed SSCAN, we design a spectral group cross attention module (SGCAM) that employs a group convolution and a cross-attention module to fully exploit the spectral and spatial information in HSIs. In addition, the proposed SGCAM reduces the number of parameters for feature extraction, leading to great efficiency. We also propose a spatial-spectral attention block (SSAB) that consists of a spectral attention module and a spatial attention module in our work. Deep feature extraction for the group bands feature map is achieved by cascading several SSABs. The main contributions of this work can be summarized as follows.
	
	1) We propose a novel HSIs denoising framework, both qualitative and quantitative results confirm that the proposed SSCAN outperforms existing state-of-the-art methods.
	
	2) We propose a spectral group cross attention module (SGCAM) to combine several adjacent bands and exploit the correlations between bands in HSIs, which mitigates spectral distortion. 
	
	3) In order to focus on more relevant information and more important areas, we introduce a spatial-spectral attention block into the feature extraction.
	
	
	
	The remainder of this paper is organized as follows. Section \ref{s2} describes the proposed SSCAN in detail. Section \ref{s3} details the experimental and presents the results. Section \ref{s4} concludes this study.
	\vspace{-0.5cm}
	\section{Methodology}\label{s2}
	In this section, we describe our proposed spatial-spectral cross attention network (SSCAN) in detail. 
	\vspace{-0.5cm}
	\subsection{Network Architecture}
    We propose SSCAN, a novel HSI denoising network that consists of a group cross attention module and a spectral-spatial attention module. We demonstrate the network architecture of the SSCAN in Fig. \ref{fig1}. ${I_{input}}$ denotes the input HSIs with noises, and the goal of the proposed network ${H_{SSCAN}}$ is to generate the corresponding noise-free output ${I_{generate}}$. This process can be described as:
	\begin{equation}
		{I_{generate} = {H_{SSCAN}}({I_{input}})}.
	\end{equation}
	\vspace{-0.5cm}
	\begin{figure}[h]
		\includegraphics[width=9cm]{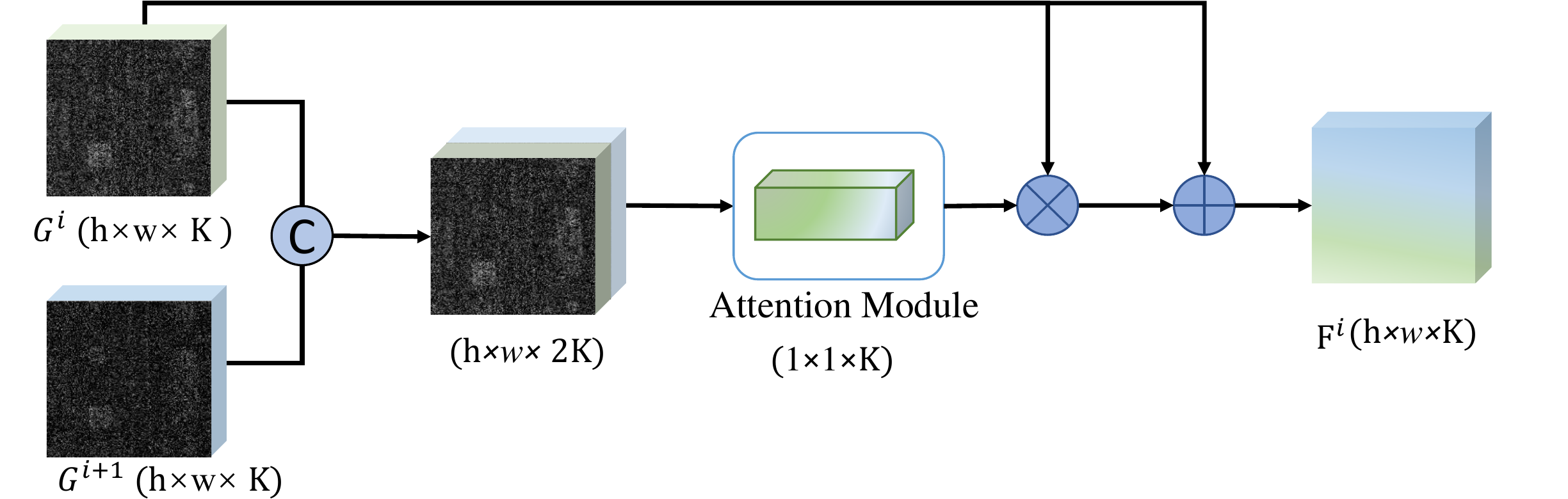}
		\vspace{-0.25cm}
		\caption{The detailed structure of Spectral Grouped Cross Attention Module (SGCAM). “+”and “×” denote element-wise addition and element-wise multiplication, respectively.\label{fig2}}
	\end{figure}
	\vspace{-1cm}
	\subsection{Spectral Grouped Cross Attention Module (SGCAM)}
	It has been proved that bands from HSIs contain rich spectral and spatial information and are highly correlated with each other\cite{jiang2020learning}. Thus, treating them separately as an ordinary gray-scale band fails to consider the interactions among different bands, while treating all bands as a whole usually leads to a large number of parameters. 

	In this study, we propose a spectral group cross attention module (Fig \ref{fig2}). First, SGCAM divide the given noisy HSI into several overlapping groups along the spectral space $G = \{{G^0},{G^1},...,{G^n}\}$, where $n$ denotes the total channel number of band groups. Each group contains $k$ bands and the adjacent band groups $G^i$,$G^{i+1}$ have $o$ bands overlapping.
	To further exploit the spectral correlation between adjacent band groups, we propose a cross-attention module. For each group ${G^i}(i<n-1)$, we incorporate its next band group ${G^{i+1}}(i<n-1)$ to obtain a feature map ${F^i}$ using cross attention module,
	\begin{equation}
	{F^i} = {H_{CAM}}({G^i},{G^{i + 1}}),
	\end{equation}
	where ${H_{CAM}}$ denotes cross attention module, $F^i$ is the feature map of $G^i$.

	In ${H_{CAM}}$, we concatenate the band group $G^i$ and $G^{i+1}$, obtain a feature map with bands of both of them, and acquire the attention mask to highlight the spatial feature in $G^i$. The attention module focuses on the relationship between channels and facilitates the automatic learning of the feature importance within different channels. The cross attention operation between adjacent band groups establishes the correlations among groups, thus leading to more efficient spatial-spectral feature extraction.

	\vspace{-0.5cm}
	\subsection{Spectral-spatial Attention Network (SSAB)}
	
	In HSI denoising, it is important to exploit the non-local spatial self-similarity and correlation among the spectral bands. Manually designed constraints from traditional algorithms are not sufficient for accurate HSI denoising. To take advantage of the spectral and spatial information in HSI, we advocate a novel feature extraction block, Spectral-spatial Attention Block (SSAB) to improve the feature extraction performance.  
	\vspace{-0.5cm}
	\begin{figure}[h]
		\centering
		\includegraphics[width=9 cm]{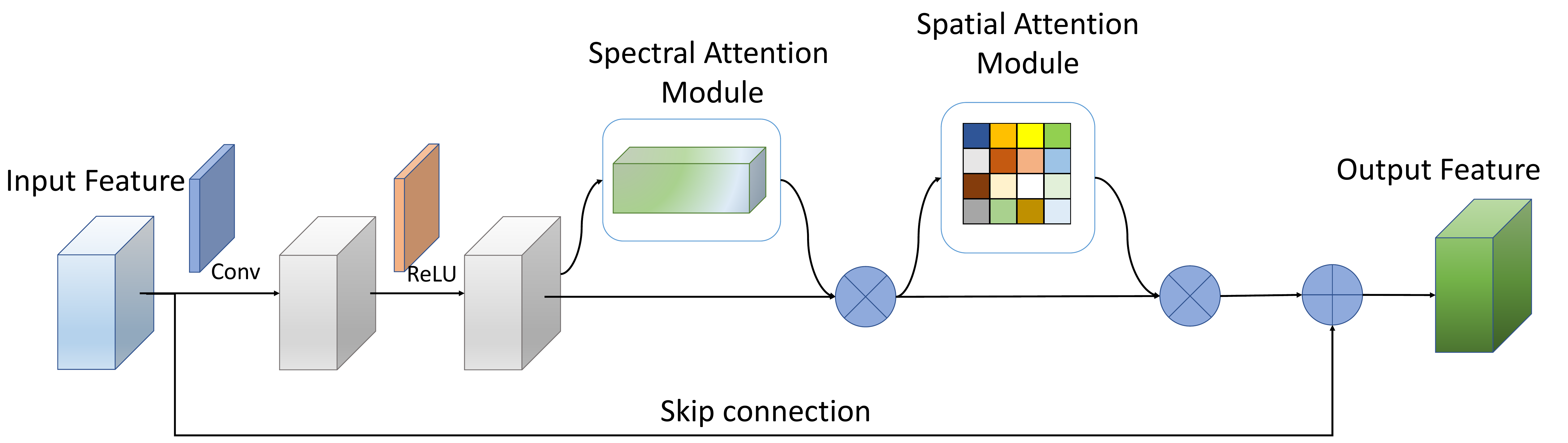}
		\caption{The detailed structure of Spectral-spatial Attention Block (SSAB)\label{fig3}}
	\end{figure}
	
	The architecture of SSAB is illustrated in Fig\ref{fig3}. SSAB cascades a spectral attention module and a spatial attention module in a sequential manner. The spectral attention module directs model’s attention to meaningful bands, while the spatial attention module directs model’s attention to meaningful regions. To achieve fast and stable training and to pass low-frequency features to the end, SSAB incorporates residual blocks.
	\begin{equation}
	{H_{SSAB}} = {H_{spea}}({H_{spca}}({F^i})) + {F^i},
	\end{equation}
	where $F^i$ refers to the $i-th$ band group feature map. ${H_{spca}}$ and ${H_{spea}}$ denote the spectral attention module and the spatial attention module, respectively. ${H_{SSAB}}$ is the SSAB operation, $+ {F^i}$ can be referred to as residual learning. 
	
	In the spectral attention module ${H_{spca}}$, we generate a spectral attention map ${M_{spca}}$ using the inter-band relationships within channels. In this module, we utilize a max-pooling operation and an average-pooling operation to aggregate the spatial information in the spatial dimension of feature map ${F^0}$, which are fed into two CNNs with shared parameters with ReLU as the activation function. The two output 1D features are then concatenated. We use a sigmoid operation to obtain the attention matrix ${F_{spca}}$, that multiplies the input feature map and result in a new feature map ${M_{spca}}$.
	
	In the spatial attention module ${M_{spea}}$, we generate a spatial attention map using the inter-spatial relationship of feature maps. First, we perform the max-pooling and average-pooling in the channel dimension to obtain two channel descriptors with ReLU as the activation function. Different from the previous ones, we futher concatenate them together. After a convolution layer with the sigmoid activation function, we obtain the weight coefficients. We further multiply it and the input feature to obtain the final output feature map.
	
	In the feature extraction of SSCAN, we cascade $n$ SSABs as a spatial-spectral attention Network (SSAN) to exploit high-frequency feature in each band group, and the output features can be formulated by:
	\begin{equation}
	F_{SSAN}^i = {H_{SSAB}^n}(H_{SSAB}^{n - 1}(...{H_{SSAB}^1}({F^i})...)) + {F^i}
	\end{equation}
	where $F_{SSAN}^i$ is the final extracted feature, and $H_{SSAB}^n$ denotes the function of the $n$-th SSAB.
	
	In global network, after extracting features from every bands groups, we concatenate them and feed them into another SSPN to get the final feature map.
	\vspace{-0.35cm}
	\subsection{Loss Function}
	 We use the Mean Squared Error (MSE) as the loss function in SSCAN. MSE is a classical loss function that has been widely applied in regression problems. It is calculated as the average of the squared differences between the denoising image and the clean image: 
	\begin{equation}
	{MSE}(\Theta ) = \frac{1}{2N}{\sum\limits_{n = 1}^N {\left\| {I_{gt}^n - {H_{SSCAN}}({I^n})} \right\|}^2 },
	\end{equation}
	where $\Theta$ is the parameter that needs to be learn in the proposed SSCAN, $I^n$ denotes the noisy HSI, $I_{gt}^n$ denotes the ground truth clean HSI, and $N$ is the number of images in a batch.
	
		\begin{table}[h]
		\centering
		\caption{Quantitative evaluation of the proposed SSCAN and other comparing methods in HSI image denoising task with four noise levels ($\sigma = 5,25,50$ and $75$). Values highlighted by \textcolor{red}{Red} indicates the best while values highlighted by the \textcolor{blue}{blue} indicates the second best performance.
			\label{tab1}}
		
		\renewcommand\arraystretch{1.2}

		\begin{tabular}{|c|c|c|c|c|c|}
			\hline
			Method & $\sigma_{n}$  & MPSNR $\uparrow$     & MSSIM $\uparrow$       & SAM $\downarrow$ &ERGAS$\downarrow$ \\ \hline \hline
			Noisy HSI &5   &34.15 &0.8927   &4.0775   &3.5680  \\  \cline{1-6} 
			LRTA \cite{renard2008denoising}  &5   &43.67 &0.9851   &1.4136   &1.0310 \\  \cline{1-6}
			NGMeet \cite{he2019non}  &5   &\textcolor{blue}{44.51} &\textcolor{blue}{0.9920}   &\textcolor{blue}{1.1581}   &\textcolor{blue}{1.1067} \\  \cline{1-6} 
			DNCNN \cite{zhang2017beyond}  &5    &38.05 &0.9624   &2.7628   &3.5680 \\  \cline{1-6} 
			Memnet \cite{tai2017memnet}  &5   &37.04   &0.9693 &3.4100      &2.4523\\  \cline{1-6}    
			HSID \cite{yuan2018hyperspectral} &5   &42.34 &0.9855    &1.5743   &1.1686 \\  \cline{1-6} 
			HSI-DeNet \cite{chang2018hsi}   &5     &37.73 & 0.9786   &2.1208   &2.2816 \\ \cline{1-6} 
			GradNet \cite{liu2020gradnet}   &5   &43.87 &0.9903    &1.2365   &0.9971 \\ \cline{1-6}
			SSCAN(Ours)  &5 &\textcolor{red}{45.32}  &\textcolor{red}{0.9930}  &\textcolor{red}{1.0549}   &\textcolor{red}{0.8469} \\ \cline{1-6} \hline \hline
			Noisy HSI &25  & 20.17   & 0.3553  &19.5171  &15.6678 \\  \cline{1-6} 
			LRTA \cite{renard2008denoising}  &25   &35.62 &0.9348   &3.2131   &2.3977 \\  \cline{1-6}
			NGMeet \cite{he2019non}  &25   &38.61 &0.9638 &2.2261   &1.6973 \\  \cline{1-6}
			DNCNN \cite{zhang2017beyond}  &25     &35.56  & 0.9653   &2.9029   &2.3700\\  \cline{1-6} 
			Memnet \cite{tai2017memnet}  &25    &31.27   &0.9246    &7.7444   &3.9550\\  \cline{1-6}    
			HSID \cite{yuan2018hyperspectral} &25  &35.34 & 0.9385&3.3599      &2.5004 \\  \cline{1-6} 
			HSI-DeNet \cite{chang2018hsi}   &25    &37.28 &0.9713 & 2.2605     &2.1239 \\ \cline{1-6} 
			GradNet \cite{liu2020gradnet}   &25   &\textcolor{blue}{39.27} &\textcolor{blue}{0.9739}   &\textcolor{blue}{2.0372}  &\textcolor{blue}{1.5958}  \\ \cline{1-6}
			SSCAN(Ours)  &25 &\textcolor{red}{39.78}   &\textcolor{red}{0.9770}   &\textcolor{red}{1.9479}   &\textcolor{red}{1.4931} \\ \cline{1-6} \hline \hline
			Noisy HSI &50  & 14.15  &0.1495    &35.0295   &31.3355  \\  \cline{1-6} 
			LRTA \cite{renard2008denoising}  &50   &32.19 &0.8693   &4.6755  &3.5088 \\  \cline{1-6}
			NGMeet \cite{he2019non}  &50   &\textcolor{blue}{36.37} &0.9513   &2.7739   &\textcolor{blue}{2.1868} \\  \cline{1-6}
			DNCNN \cite{zhang2017beyond}  &50      & 34.95  &0.9404 &4.0269   &3.4534\\  \cline{1-6} 
			Memnet \cite{tai2017memnet}  &50   &30.11  &0.8800 &7.0705      &4.3843\\  \cline{1-6}    
			HSID \cite{yuan2018hyperspectral} &50    &31.75 &0.8770 &4.8834      &3.6485 \\  \cline{1-6} 
			HSI-DeNet \cite{chang2018hsi}   &50   &35.81 &\textcolor{red}{0.9569} &\textcolor{blue}{2.6711}   &2.4566 \\ \cline{1-6} 
			GradNet \cite{liu2020gradnet}   &50   &35.99 &0.9471 & 2.9066  &2.2764 \\ \cline{1-6}
			SSCAN(Ours)  &50  &\textcolor{red}{36.73} &\textcolor{blue}{0.9564} &\textcolor{red}{2.6162}   &\textcolor{red}{2.0739}     \\ \cline{1-6} \hline \hline
			Noisy HSI &75  &10.63   &0.0779 & 46.1527      &47.0033  \\  \cline{1-6} 
			LRTA \cite{renard2008denoising}  &75   &30.12 &0.8077  &5.8103   &4.4231 \\  \cline{1-6}
			NGMeet \cite{he2019non}  &75   &34.29 &0.9273   &3.4683   &2.7821 \\  \cline{1-6}
			DNCNN \cite{zhang2017beyond}  &75     &34.01   &0.9222 &3.8142   & 3.7451  \\  \cline{1-6}
			Memnet \cite{tai2017memnet}  &75      &28.34   &0.8280 &7.2952   &5.4204   \\  \cline{1-6}  
			HSID \cite{yuan2018hyperspectral} &75    &30.32 &0.8276 &5.8687   &4.3699    \\  \cline{1-6} 
			HSI-DeNet \cite{chang2018hsi}   &75  &33.95 &\textcolor{blue}{0.9322} &\textcolor{blue}{3.3979}  &3.4858   \\ \cline{1-6} 
			GradNet \cite{liu2020gradnet}   &75    &\textcolor{blue}{34.49} &0.9255 &{3.4288}      &\textcolor{blue}{2.6565} \\ \cline{1-6}
			SSCAN(Ours)  &75   &\textcolor{red}{35.36}  &\textcolor{red}{0.9350} &\textcolor{red}{2.9655}   &\textcolor{red}{2.4178}    \\ \cline{1-6} \hline \hline
			
		\end{tabular}
	\end{table}
	\vspace{-0.25cm}
	\section{Experiments}\label{s3}
	In this section, we validate the effectiveness of SSCAN on a public hyperspectral image dataset. The proposed method is compared with several existing state-of-the-art methods of traditional HSI denoising, including low-rank tensor approximation (LRTA) \cite{renard2008denoising} and NG-Meet \cite{he2019non}. We also choose five deep learning methods for comparison: DNCNN \cite{zhang2017beyond}, Memnet \cite{tai2017memnet}, HSID \cite{yuan2018hyperspectral}, HSI-DeNet \cite{chang2018hsi}, and GradNet \cite{liu2020gradnet}.

	The Washington DC Mall image (with a total of 210 bands) obtained by the Hyperspectral Digital Imagery Collection Experiment airborne sensor in August 1995 is utilized in our experiment. Due to the noises generated during the data acquisition, the number of available bands is 191. The experimental hyperspectral image has a size of 1,208 $\times$ 303 pixels. We split it into two parts: 1,080 $\times$ 303 $\times$ 191 as the training set and 200 $\times$ 200 $\times$ 191 as the testing set.  We further scale values in each band to a range of [0,1].
	Noises are added to HSIs following a certain procedure. For every band in HSI, the noise intensity is confined to a Gaussian distribution, with four different settings of 5, 25, 50, and 75.
    \vspace{-0.5cm}
	\subsection{Evaluation Metrics}
	To quantitatively evaluate the performance of different denoising methods, we adopt the following evaluation metrics in our simulated experiments: mean peak signal-to-noise ratio (MPSNR) \cite{singh2014quality}, mean structural similarity index (MSSIM) \cite{wang2004image}, spectral angle mapper (SAM) \cite{yuhas1992discrimination} and Erreur Relative Globale Adimensionnelle de Synthèse (ERGAS) \cite{veganzones2015hyperspectral}.
	\vspace{-0.5cm}
	\subsection{Implementation Details}
	In our experiments, the training and testing of the proposed SSCAN, as well as other comparative methods, are conducted under the Pytorch \footnote{\url{https://pytorch.org/}} environment. All experiments run on a PC with two NVIDIA 2080Ti GPUs. We choose the ADAM \cite{kingma2014adam} algorithm as the optimizer with $\beta_1 = 0.9$, $\beta_2 = 0.999$, and $\epsilon = 1e-8$. The initial learning rate is $1e-4$, which decays by a factor of 10 when the epoch achieves 50. The number of each band group $K$ is set to 4, the overlapping parameter $o$ is set to 2, and the number of SSAB $n$ is set to 10.
	\vspace{-0.15cm}
	\begin{figure}[h]
		\includegraphics[width=9cm]{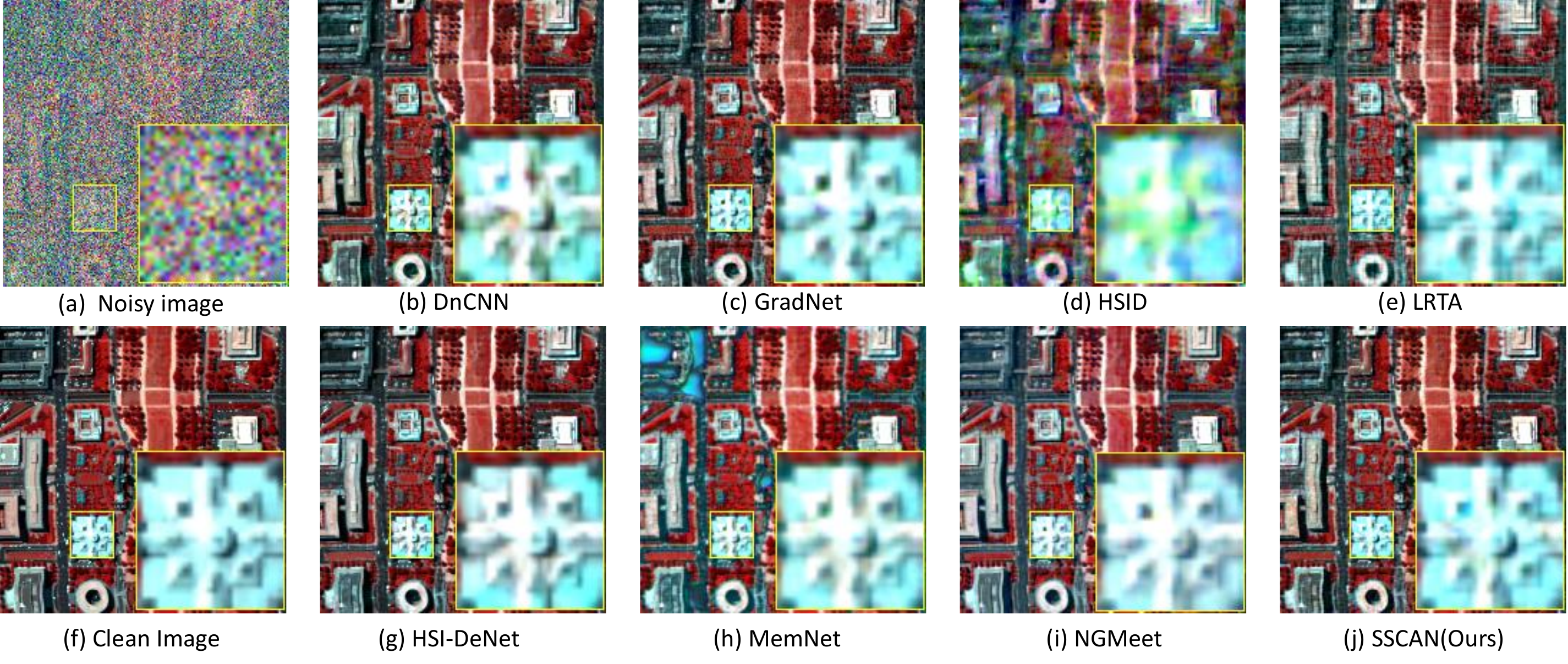}
		\caption{Result for the Washington dc Mall with $\sigma_{n}$= 50.(a) noisy image with bands (57,27,17), (b) DNCNN, (c) GradNet, (d) HSID, (e) clean image, (f) LRTA, (g) HSI-DeNet, (h) MemNet, (i) NGMeet, (j) SSCAN(ours). \label{fig4}}
	\end{figure}
	\subsection{Model Performance}
	Table \ref{tab1} presents the quantitative evaluation of the proposed SSCAN and other comparing methods in the HSI image denoising task with different noise levels ($\sigma_n$ = 5, 25, 50, and 75). We observe that the MPSNR values of the proposed method are the highest at most noise levels. NGMeet achieves the second-best performance at $\sigma=25$ and $\sigma = 75$. Similarly, the proposed SSCAN obtains the best performance in MSSIM at most noise levels, except $\sigma = 50$ (HSI-DeNet\cite{chang2018hsi} outperforms SSCAN in terms of MSSIM). As for the other two evaluation metrics, i.e., ERGAS and SAM, that reflect the denoising effect in the spectral domain, the proposed SSCAN achieves the best performance at all noise levels.
	
\begin{figure}[h]
		\includegraphics[width=9 cm]{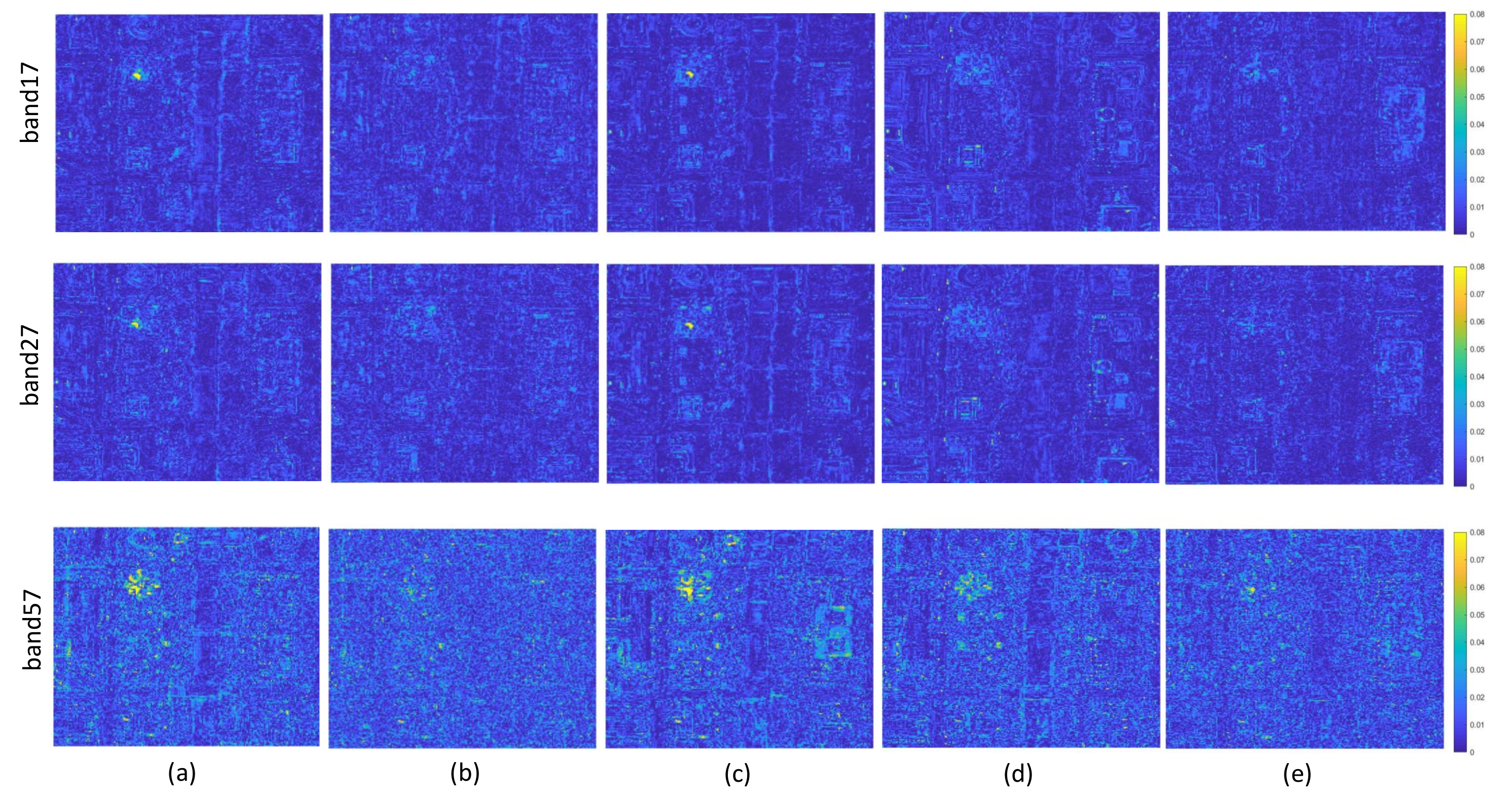}
		\centering
		\caption{The error maps the Washington dc Mall with $\sigma_{n}$= 50 with spectral bands (57,27,17) (a) DNCNN, (b) GradNet, (c)HSI-DeNet, (d) MemNet, (e) NGMeet, (f) SSCAN(ours). \label{fig5}}
	\end{figure}

	We also evaluate model performance between SSCAN and other competing methods in HIS denoising in a qualitative manner (Fig.\ref{fig4}). We notice that noises still exist in the results of LRTA \cite{renard2008denoising}, MemNet \cite{tai2017memnet}, and HSID \cite{yuan2018hyperspectral}. In comparison, the denoised image from SSCAN is considerably closer to the original image in terms of image details and color representation. We also present the error maps for the result of DNCNN \cite{zhang2017beyond}, GradNet \cite{liu2020gradnet}, HSI-DeNet \cite{chang2018hsi}, NGMeet \cite{tai2017memnet}, and SSCAN in Fig.\ref{fig5}, aiming to better evaluate the denoising performance of these methods. We observe that no notable contours nor light-colored regions in the proposed SSCAN compared to other competing algorithms, indicating the great denoising performance of the proposed method.
	
	\section{Conclusion}\label{s4}
	In this study, we propose a spatial-spectral cross attention network (SSCAN) for hyperspectral image denoising. The proposed SSCAN combines group convolution and attention modules to achieve state-of-the-art denoising performance. To take full advantage of the correlation between adjacent spectral bands while reducing the parameters of feature extraction, we design a spectral group cross attention module (SGCAM). To exploit the spatial and spectral information in hyperspectral images in an effective manner, we design a spatial-spectral attention block (SSAB). In addition, we adopt residual learning operations to facilitate fast and stable converge. The experimental results indicate that the proposed SSCAN achieves better denoising performance compared with other selected competing methods qualitatively and quantitatively.

	\bibliographystyle{IEEEtran}
	\bibliography{IEEEabrv,ac}

\begin{thebibliography}{10}
\providecommand{\url}[1]{#1}
\csname url@samestyle\endcsname
\providecommand{\newblock}{\relax}
\providecommand{\bibinfo}[2]{#2}
\providecommand{\BIBentrySTDinterwordspacing}{\spaceskip=0pt\relax}
\providecommand{\BIBentryALTinterwordstretchfactor}{4}
\providecommand{\BIBentryALTinterwordspacing}{\spaceskip=\fontdimen2\font plus
\BIBentryALTinterwordstretchfactor\fontdimen3\font minus
  \fontdimen4\font\relax}
\providecommand{\BIBforeignlanguage}[2]{{%
\expandafter\ifx\csname l@#1\endcsname\relax
\typeout{** WARNING: IEEEtran.bst: No hyphenation pattern has been}%
\typeout{** loaded for the language `#1'. Using the pattern for}%
\typeout{** the default language instead.}%
\else
\language=\csname l@#1\endcsname
\fi
#2}}
\providecommand{\BIBdecl}{\relax}
\BIBdecl

\bibitem{kruse2003comparison}
F.~A. Kruse, J.~W. Boardman, and J.~F. Huntington, ``Comparison of airborne
  hyperspectral data and eo-1 hyperion for mineral mapping,'' \emph{IEEE
  transactions on Geoscience and Remote Sensing}, vol.~41, no.~6, pp.
  1388--1400, 2003.

\bibitem{yang2017learning}
J.~Yang, Y.-Q. Zhao, and J.~C.-W. Chan, ``Learning and transferring deep joint
  spectral--spatial features for hyperspectral classification,'' \emph{IEEE
  Transactions on Geoscience and Remote Sensing}, vol.~55, no.~8, pp.
  4729--4742, 2017.

\bibitem{bioucas2012hyperspectral}
J.~M. Bioucas-Dias, A.~Plaza, N.~Dobigeon, M.~Parente, Q.~Du, P.~Gader, and
  J.~Chanussot, ``Hyperspectral unmixing overview: Geometrical, statistical,
  and sparse regression-based approaches,'' \emph{IEEE journal of selected
  topics in applied earth observations and remote sensing}, vol.~5, no.~2, pp.
  354--379, 2012.

\bibitem{ghamisi2017advances}
P.~Ghamisi, N.~Yokoya, J.~Li, W.~Liao, S.~Liu, J.~Plaza, B.~Rasti, and
  A.~Plaza, ``Advances in hyperspectral image and signal processing: A
  comprehensive overview of the state of the art,'' \emph{IEEE Geoscience and
  Remote Sensing Magazine}, vol.~5, no.~4, pp. 37--78, 2017.

\bibitem{2018Noise}
R.~Behnood, S.~Paul, G.~Pedram, L.~Giorgio, and C.~Jocelyn, ``Noise reduction
  in hyperspectral imagery: Overview and application,'' \emph{Remote Sensing},
  vol.~10, no.~3, p. 482, 2018.

\bibitem{2018Multifeature}
H.~Su, B.~Zhao, Q.~Du, P.~Du, and Z.~Xue, ``Multifeature dictionary learning
  for collaborative representation classification of hyperspectral imagery,''
  \emph{IEEE Transactions on Geoence and Remote Sensing}, pp. 1--18, 2018.

\bibitem{rasti2018noise}
B.~Rasti, P.~Scheunders, P.~Ghamisi, G.~Licciardi, and J.~Chanussot, ``Noise
  reduction in hyperspectral imagery: Overview and application,'' \emph{Remote
  Sensing}, vol.~10, no.~3, p. 482, 2018.

\bibitem{dabov2007image}
K.~Dabov, A.~Foi, V.~Katkovnik, and K.~Egiazarian, ``Image denoising by sparse
  3-d transform-domain collaborative filtering,'' \emph{IEEE Transactions on
  image processing}, vol.~16, no.~8, pp. 2080--2095, 2007.

\bibitem{rudin1992nonlinear}
L.~I. Rudin, S.~Osher, and E.~Fatemi, ``Nonlinear total variation based noise
  removal algorithms,'' \emph{Physica D: nonlinear phenomena}, vol.~60, no.
  1-4, pp. 259--268, 1992.

\bibitem{buades2005non}
A.~Buades, B.~Coll, and J.-M. Morel, ``A non-local algorithm for image
  denoising,'' in \emph{Computer Vision and Pattern Recognition}, vol.~2.\hskip
  1em plus 0.5em minus 0.4em\relax IEEE, 2005, pp. 60--65.

\bibitem{maggioni2012nonlocal}
M.~Maggioni, V.~Katkovnik, K.~Egiazarian, and A.~Foi, ``Nonlocal
  transform-domain filter for volumetric data denoising and reconstruction,''
  \emph{IEEE transactions on image processing}, vol.~22, no.~1, pp. 119--133,
  2012.

\bibitem{chen2014denoising}
G.~Chen, T.~D. Bui, K.~G. Quach, and S.-E. Qian, ``Denoising hyperspectral
  imagery using principal component analysis and block-matching 4d filtering,''
  \emph{Canadian Journal of Remote Sensing}, vol.~40, no.~1, pp. 60--66, 2014.

\bibitem{renard2008denoising}
N.~Renard, S.~Bourennane, and J.~Blanc-Talon, ``Denoising and dimensionality
  reduction using multilinear tools for hyperspectral images,'' \emph{IEEE
  Geoscience and Remote Sensing Letters}, vol.~5, no.~2, pp. 138--142, 2008.

\bibitem{zhang2013hyperspectral}
H.~Zhang, W.~He, L.~Zhang, H.~Shen, and Q.~Yuan, ``Hyperspectral image
  restoration using low-rank matrix recovery,'' \emph{IEEE transactions on
  geoscience and remote sensing}, vol.~52, no.~8, pp. 4729--4743, 2013.

\bibitem{he2019non}
W.~He, Q.~Yao, C.~Li, N.~Yokoya, and Q.~Zhao, ``Non-local meets global: An
  integrated paradigm for hyperspectral denoising,'' in \emph{Computer Vision
  and Pattern Recognition}, 2019, pp. 6868--6877.

\bibitem{zhang2017beyond}
K.~Zhang, W.~Zuo, Y.~Chen, D.~Meng, and L.~Zhang, ``Beyond a gaussian denoiser:
  Residual learning of deep cnn for image denoising,'' \emph{IEEE transactions
  on image processing}, vol.~26, no.~7, pp. 3142--3155, 2017.

\bibitem{anwar2019real}
S.~Anwar and N.~Barnes, ``Real image denoising with feature attention,'' in
  \emph{International Conference on Computer Vision}, 2019, pp. 3155--3164.

\bibitem{tai2017memnet}
Y.~Tai, J.~Yang, X.~Liu, and C.~Xu, ``Memnet: A persistent memory network for
  image restoration,'' in \emph{Proceedings of the IEEE international
  conference on computer vision}, 2017, pp. 4539--4547.

\bibitem{chen2016trainable}
Y.~Chen and T.~Pock, ``Trainable nonlinear reaction diffusion: A flexible
  framework for fast and effective image restoration,'' \emph{IEEE transactions
  on pattern analysis and machine intelligence}, vol.~39, no.~6, pp.
  1256--1272, 2016.

\bibitem{xie2017hyperspectral}
W.~Xie and Y.~Li, ``Hyperspectral imagery denoising by deep learning with
  trainable nonlinearity function,'' \emph{IEEE Geoscience and Remote Sensing
  Letters}, vol.~14, no.~11, pp. 1963--1967, 2017.

\bibitem{yuan2018hyperspectral}
Q.~Yuan, Q.~Zhang, J.~Li, H.~Shen, and L.~Zhang, ``Hyperspectral image
  denoising employing a spatial--spectral deep residual convolutional neural
  network,'' \emph{IEEE Transactions on Geoscience and Remote Sensing},
  vol.~57, no.~2, pp. 1205--1218, 2018.

\bibitem{dong2019deep}
W.~Dong, H.~Wang, F.~Wu, G.~Shi, and X.~Li, ``Deep spatial--spectral
  representation learning for hyperspectral image denoising,'' \emph{IEEE
  Transactions on Computational Imaging}, vol.~5, no.~4, pp. 635--648, 2019.

\bibitem{jiang2020learning}
J.~Jiang, H.~Sun, X.~Liu, and J.~Ma, ``Learning spatial-spectral prior for
  super-resolution of hyperspectral imagery,'' \emph{IEEE Transactions on
  Computational Imaging}, vol.~6, pp. 1082--1096, 2020.

\bibitem{chang2018hsi}
Y.~Chang, L.~Yan, H.~Fang, S.~Zhong, and W.~Liao, ``Hsi-denet: Hyperspectral
  image restoration via convolutional neural network,'' \emph{IEEE Transactions
  on Geoscience and Remote Sensing}, vol.~57, no.~2, pp. 667--682, 2018.

\bibitem{liu2020gradnet}
Y.~Liu, S.~Anwar, L.~Zheng, and Q.~Tian, ``Gradnet image denoising,'' in
  \emph{Computer Vision and Pattern Recognition Workshops}, 2020, pp. 508--509.

\bibitem{singh2014quality}
A.~K. Singh, H.~Kumar, G.~Kadambi, J.~Kishore, J.~Shuttleworth, and
  J.~Manikandan, ``Quality metrics evaluation of hyperspectral images,''
  \emph{International Archives of the Photogrammetry, Remote Sensing and
  Spatial Information Sciences}, vol.~8, 2014.

\bibitem{wang2004image}
Z.~Wang, A.~C. Bovik, H.~R. Sheikh, and E.~P. Simoncelli, ``Image quality
  assessment: from error visibility to structural similarity,'' \emph{IEEE
  transactions on image processing}, vol.~13, no.~4, pp. 600--612, 2004.

\bibitem{yuhas1992discrimination}
R.~H. Yuhas, A.~F. Goetz, and J.~W. Boardman, ``Discrimination among semi-arid
  landscape endmembers using the spectral angle mapper (sam) algorithm,'' in
  \emph{Proc. Summaries 3rd Annu. JPL Airborne Geosci. Workshop}, vol.~1, 1992,
  pp. 147--149.

\bibitem{veganzones2015hyperspectral}
M.~A. Veganzones, M.~Simoes, G.~Licciardi, N.~Yokoya, J.~M. Bioucas-Dias, and
  J.~Chanussot, ``Hyperspectral super-resolution of locally low rank images
  from complementary multisource data,'' \emph{IEEE Transactions on Image
  Processing}, vol.~25, no.~1, pp. 274--288, 2015.

\bibitem{kingma2014adam}
D.~P. Kingma and J.~Ba, ``Adam: A method for stochastic optimization,''
  \emph{arXiv preprint arXiv:1412.6980}, 2014.

\end{thebibliography}
	
\end{document}